%% file: preprint_body.tex
\documentclass[runningheads]{svmult}
\usepackage{graphicx}
\usepackage{color} 
\usepackage{amsmath}
\usepackage{amssymb}
\usepackage{multicol}
\usepackage{makeidx}   

\newcommand{\HAind}[1]{#1\index{#1}}
\newcommand{\HAbegIND}[1]{\index{#1|(}}
\newcommand{\HAendIND}[1]{\index{#1|)}}

\input{qiv_definitions}
\newlength{\figurewidth}

\begin{document}


\setcounter{page}{-1}
\thispagestyle{empty}
\begin{center}
\bfseries\Large%
\par\vspace*{0.3\textheight}\par
Quantum Communication and Decoherence
\par\vspace*{2\baselineskip}\par\large\normalfont%
Hans Aschauer and Hans J. Briegel
\par\vfill\par\normalsize%
To be published in ``Coherent Evolution in Noisy Environments'',\\
Lecture Notes in Physics, \texttt{http://link.springer.de/series/lnpp/}\\
\copyright\ Springer Verlag, Berlin-Heidelberg-New York\\
\end{center}
\cleardoublepage

\title{Quantum Communication and Decoherence}
\author {Hans Aschauer\inst{1}
\and Hans J. Briegel\inst{1}}
\institute{
  Theoretische Physik, Ludwig-Maximilians-Universit\"at, \\
  Theresienstr. 37, D-80333 M\"unchen}

\maketitle              

\input{introduction_ed.tex}
\input{hans_b_ed.tex}
\input{purification_ed.tex}
\input{cryptography_ed.tex}
\input{factorization_ed.tex}

\input{acknowledgement.tex}


\input{references}
\end{document}

%% file: qiv_definitions.tex
\def\Ket#1{\left|#1\right\rangle} 
\def\Bra#1{\left\langle#1\right|}
\def\KetBra#1#2{\Ket{#1}\!\Bra{#2}} 

\def\Proj#1{\KetBra{#1}{#1}}
\def\ProjInd#1#2{\Ket{#1}_{\!#2}\!\Bra{#1}}  
\def\Eins{\mathbb{I}} 

\def\ie{i.\,e.\ }

\def\CCdots{\cdots}
\def\CLdots{\ldots}  
\def\BQuad{}

\DeclareMathOperator{\tr}{tr}


%% file: introduction_ed.tex
\section{Introduction}

The problem of decoherence is an integral part of the theory of
quantum computation and communication. The potential of a quantum
computer lies in its ability to process information in the form of a
coherent superposition of quantum mechanical states. Quantum
algorithms such as Shor's algorithm \cite{shor:97} make use of the
interference of different ``computational paths'', which can strongly
enhance their efficiency compared to classical algorithms. Because
quantum coherence and interference play a central role in a quantum
computer, {\em de}coherence is a major threat to its proper
functioning.

A similar situation prevails in quantum communication. 
The central problem of quantum communication is how to faithfully transmit
unknown quantum states through a noisy quantum channel. While quantum 
information is sent through a channel such as an optical fiber, the 
carriers of the information (e.g.\ photons) interact with the channel 
and get entangled with its many degrees of freedom, which gives rise 
to the phenomenon of decoherence on the state space of the information 
carriers. An intially pure state becomes a mixed state when it leaves 
the channel. For quantum communication purposes, it is however essential 
that the transmitted qubits retain their genuine quantum properties, for 
example in form of an entanglement with qubits on the other side of the 
channel.

To deal with the problem of decoherence, two methods have been
developed, known as {\em \HAind{quantum error correction}}
\cite{shor-code:95,calderbank-shor:96,steane-PRL:95} and {\em
  \HAind{entanglement purification}}
\cite{BBPSSW:96,bennett-divincenzo-et-al:96,DEJMPS:96}, respectively.
In quantum error correction, which will be discussed in the next
section, quantum information is encoded in the joint state of several
two-state particles, forming a so-called quantum error correcting
code, before it is sent through the channel.  By measuring certain
joint observables of the particles (the so-called stabilizer of the
code), it is thereby possible to ``reset'' the state of the
information carriers after a given time, by projecting their joint
state onto certain subspaces of their Hilbert space without destroying
the coherence of the encoded information.  Even though quantum error
correction can be used, in principle, to send quantum information
through a noisy channel, it has been primarily developed to stabilize
a quantum computer against the effect of decoherence. Entanglement
purification, on the other hand, together with the method of
\HAind{teleportation} \cite{bennett-brassard-et-al:93}, is a powerful tool
that is particularly suitable for quantum communication. The idea of
entanglement purification is to ``distill'' from an ensemble of
low-fidelity Einstein-Podolsky-Rosen (EPR\index{EPR pairs}) \cite{EPR:35} pairs, which
have been distributed through some noisy channel, a smaller ensemble
of high-fidelity EPR pairs which may then be used for faithful
teleportation \cite{BBPSSW:96} or for quantum cryptography
\cite{BB84,ekert:91}.  This distillation process requires certain
unitary operations and measurements to be performed on the qubits at
each side of the channel, and a process of postselection, which also
requires classical communication between the parties.

Both methods, quantum error correction and entanglement purification, fight 
decoherence by a process of {\em controlled \HAind{disentanglement}} of the 
information carriers from the quantum channel. This process involves 
the action of some apparatus that is used to transform and measure the 
state of the particles, for example via tunable interactions of the 
particles with each other and with external fields. 
Real apparatuses are themselves sources of noise, 
which 
complicates the 
situation considerably. From a general perspective, the apparatuses used by 
Alice and Bob must themselves be considered as part of the noisy communication
channel.  Under realistic circumstances, the information carriers will thus 
{\em always} become entangled with other degrees of freedom and therefore 
suffer from a certain amount of decoherence. The question is therefore not 
whether decoherence can be avoided at all, but whether its influence can be 
kept on a tolerable level. 

What ``tolerable'' means depends on the context. In quantum
computation\index{quantum computation}, for example, the effect of decoherence may be tolerable
as long as the fidelity of the output of a quantum algorithm is above
a certain value, allowing one to extract the desired result with the
corresponding probability. In \HAind{quantum cryptography} the effect of the
channel cannot, in principle, be distinguished from an intelligent
third party who manipulates the transmitted quantum systems to gain
information about their state. All noise of a channel is therefore
attributed --- this is the pessimistic attitude of the cryptologist
--- to an adversary.  Decoherence is thereby considered due to
entanglement of the information carriers with degrees of freedom
controlled by an adversary.  As we will show in the later part of this
review, the {\em security} of quantum cryptography is in fact closely
connected to the {\em disentanglement} of the degrees of freedom of
the information carriers, on one side, and the channel, on the other
side.  Even though we cannot avoid all residual entanglement with the
channel, we can distinguish between residual entanglement with the
apparatus, which is harmless, and residual entanglement with the part
of the channel accessible to an eavesdropper, which is potentially
harmful.

In the following, we will give a brief introduction to the methods of
quantum error correction and entanglement purification, and to the
basic protocols of quantum cryptography.\footnote{For a more
  comprehensive introduction into these fields of quantum information
  theory, see, for example, Ref.~\cite{springer:00}.} We will then
discuss a recent security proof \cite{aschauer_prl} for
entanglement-based quantum communication through noisy channels, which
explicitly takes into account the role of noisy apparatus. We try to
pay particular attention to conceptual issues but skip some of the
technical details, which can be found in the literature.


%% file: hans_b_ed.tex
\section{Quantum Error Correction}
\label{sec:quantum_error_correction}
\HAbegIND{quantum error correction}

\begin{figure}[tbp]
  \includegraphics[width=\figurewidth]{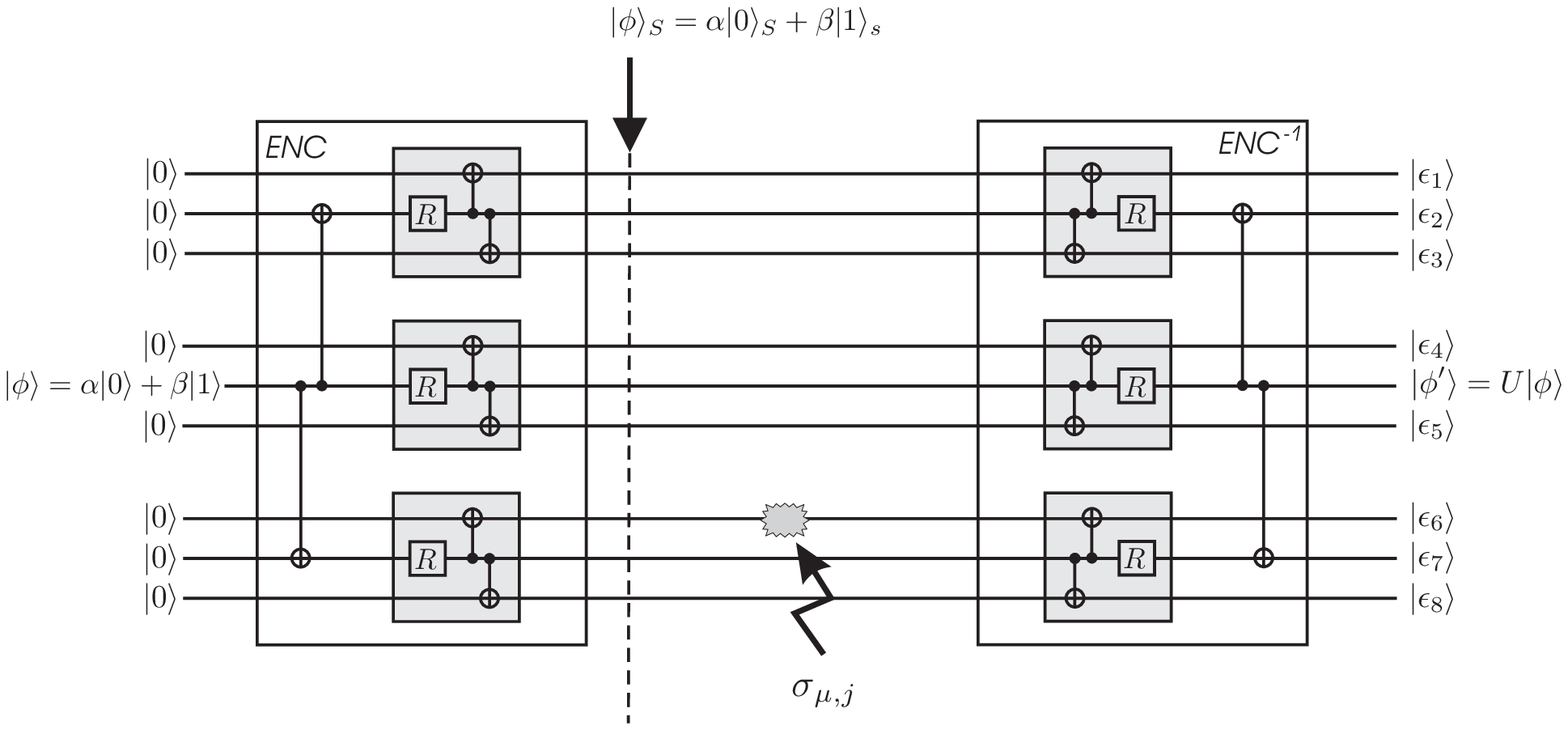}
  \caption[Quantum logic network of the Shor code \index{quantum
    codes!Shor code} and quantum error correction.]{Quantum logic
    network of the Shor code and quantum error correction.  A ``random
    rotation'' $\sigma_{\mu,j}$ on qubit $j$ in the encoded state
    translates into a certain ``\HAind{error syndrome}''
    $\epsilon_1,\dots,\epsilon_8$ and a corresponding unitary
    operation $U=U(\epsilon_1,\dots,\epsilon_8)$ on the central qubit
    (see text).  The network uses the Hadamard-Rotation $R_j = {1}/
    {\sqrt{2}}(\sigma_{x,j}+\sigma_{z,j})$ and the \HAind{CNOT}
    gate (%
    \raisebox{-0.5ex}{\includegraphics[width=2ex]{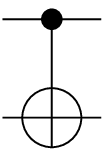}}=
    \(\mathrm{CNOT}_{i,j} =
    \frac{1+\sigma_{z,i}}{2} + \frac{1-\sigma_{z,i}}{2} \sigma_{x,j}\)).}
  \label{FIGqecc}
\end{figure}

Quantum mechanical entanglement is exploited in quantum algorithms and
in many protocols for quantum communication such as teleportation or
entang\-lement-based quantum key distribution. It also plays a
fundamental role in quantum error correction, where the coding
operations are themselves simple quantum algorithms. Let us illustrate
the basic principles at the example of the first quantum error
correcting code \index{quantum codes}found by Peter Shor in 1995
\cite{shor-code:95}\index{quantum codes!Shor code}.  To
protect quantum information that is represented by the state of a
particle (central qubit in Fig.~\ref{FIGqecc}) against decoherence,
the information is first distributed or delocalized over several
particles.  In Fig.~\ref{FIGqecc} this is done with the help of the
network $ENC$, which realizes the following mapping:
\begin{equation}
  ENC:\; (\alpha |0\rangle+\beta |1\rangle)|0\rangle|0\rangle \cdots |0\rangle
            \longmapsto \alpha |0\rangle_S + \beta |1\rangle_S
\end{equation}
in which the states
\begin{eqnarray}
 |0\rangle_S = 2^{-3/2}(|000\rangle + |111\rangle) (|000\rangle + |111\rangle) 
                                     (|000\rangle + |111\rangle) \nonumber\\
 |1\rangle_S = 2^{-3/2}(|000\rangle - |111\rangle) (|000\rangle - |111\rangle) 
                                     (|000\rangle - |111\rangle)\,. 
\end{eqnarray} 
denote the so-called {\em code words} of the (9-bit) Shor code
\index{quantum codes!Shor code}.  The encoding transformation thus
corresponds to an embedding ${\cal H} \ni |\phi\rangle = \alpha
|0\rangle+\beta |1\rangle \longmapsto \alpha |0\rangle_S + \beta
|1\rangle_S = |\phi\rangle_S \in {\cal H}_S \subset {\cal H}^{\otimes
  9} $ of the two-dimensional Hilbert space ${\cal H} \simeq
\mathbb{C}^2$ of the central qubit into the higher-dimensional Hilbert
space of all 9 qubits. After the transformation the quantum
information lies in a two-dimensional subspace ${\cal H}_S$ of a $2^9$
dimensional Hilbert space.  The code words $|0\rangle_S$ and
$|1\rangle_S$ are tensor products of entangled three-qubit states of
the form $|000\rangle \pm |111\rangle$, the so-called
Greenberger-Horne-Zeilinger (GHZ) states \cite{ghz:89}, which play a
prominent role for the interpretation of quantum mechanics.
\cite{ghz:89,mermin:90}. One can easily check that after the encoding
(see dotted line in Fig.~\ref{FIGqecc}) the reduced density operator
of each of the qubits is totally mixed; that is, the {\em individual}
state of the particles carries no information about
$|\phi\rangle$.\footnote{Quantum error correcting codes are indeed
  constructed in such a way that the state of individual qubits in a
  codeword becomes completely undetermined. As was shown by DiVincenzo
  and Peres \cite{divincenzo-peres:97}, the codewords satisfy
  generalized Mermin relations \cite{mermin:90} that exclude the
  possibility of consistently assigning a predetermined value to
  complementary observables of each qubit. From the measurement of an
  individual qubit one can thus not gain any information about
  $|\phi\rangle$. In the positive sense this means that an
  uncontrolled interaction of the environment with one of the qubits
  does not (necessarily) lead to an irreversible {\em loss} of
  information.}
 
For simplicity let us consider an error model where random rotations
are applied to the individual qubits with a certain ``error rate''.
This model is more general than it appears to be at first sight but it
needs a justification to which we shall return below. Suppose that,
after the encoding circuit of Fig.~\ref{FIGqecc}, one of the four
Pauli-Rotations $\sigma_{\mu,j}$ ($\mu=0,1,2,3$) is applied to one of
the nine qubits, where both $j$ and $\mu $ are random and unknown to
us.  The question is this: Can we still extract $|\phi\rangle$ from
the joint state of the particles? If such a random rotation is applied
to the particle {\em before} the encoding, it is clear that the
information is lost, for
$\frac{1}{4}\sum_{\mu}\sigma_{\mu}|\phi\rangle\langle\phi|\sigma_{\mu}
=\frac{1}{2}\mathbb{I}$.  If it is applied to one of the particles of
the encoded state, however, the information can still be rescued from
the joint state of all 9 particles. A possibility to do this is shown
in Fig.~\ref{FIGqecc}.  There the inverse network of $ENC$ is applied
to the code which transforms the corrupt state
$\sigma_{\mu,j}|\phi\rangle_S$, for arbitrary $\sigma_{\mu,j}$, back
to a product state:
\begin{equation}
  ENC^{-1}:\; \sigma_{\mu,j}|\phi\rangle_S \longmapsto |\phi'\rangle |\epsilon_1\rangle
  |\epsilon_2\rangle \dots |\epsilon_8\rangle\,.
\end{equation}
After this decoding transformation, the state of the neighboring
qubits carries the so-called {\em \HAind{error syndrome}} $\epsilon_1 \dots
\epsilon_8$.  The central qubit is in state $|\phi'\rangle =
U(\epsilon_1 \dots \epsilon_8) |\phi\rangle$, where the unitary
transformation $U(\epsilon_1 \dots \epsilon_8) \in \{\mathbb{I},
\sigma_x, \sigma_y,\sigma_z \}$, is uniquely determined by the error
syndrome. 

By reading off the error syndrome, i.e. measuring the state
of all neighboring qubits, and subsequently applying the correction
operation $U^{-1}(\epsilon_1 \dots \epsilon_8) $, the central qubit is
transformed back to its initial state. Please note that the central
qubit remains unmeasured, and no information about the state
$|\phi\rangle$ is obtained at any step of the protocol.  By iteration
of the sequence {\em decoding $\rightarrow$ syndrome measurement $\&$
  correction $\rightarrow$ encoding} \cite{laflamme:96} an unknown
quantum state can thus be protected against decoherence over a time
significantly longer than the decoherence time.

The effect of the random rotations $\sigma_{\mu,j}$ is to map the code
space ${\cal H}_S$ to a set of orthogonal error spaces
$\sigma_{\mu,j}{\cal H}_S \bot {\cal H}_S$.  The images of the code
words thereby satisfy the following orthogonality relations $_S\langle
0 | \sigma_{\mu,j}\sigma_{\nu,k}|1\rangle_S = 0$ and $_S\langle 0 |
\sigma_{\mu,j}\sigma_{\nu,k}|0\rangle_S = \langle 1 |
\sigma_{\mu,j}\sigma_{\nu,k}|1\rangle_S$ for all $j,k,\mu,\nu$.
Theses relations ensure
\cite{bennett-divincenzo-et-al:96,knill-laflamme:97}, that all errors
$\sigma_{\mu,j}$ can, in fact, be corrected. The Shor code was the
first quantum error correcting code found that can correct all of the
four errors (spin flip, phase flip, spin\&phase flip, identity) on any
one of the qubits. Independent of Shor, Steane \index{quantum codes!Steane code} \cite{steane-PRL:95} found a code that achieves
the same task using only 7 qubits.  Later, the theory of quantum error
correcting codes was further developed
\cite{calderbank-shor:96,steane-ROYSOC:95}, establishing in particular
the connection with classical coding theory.  A number of other codes
were found, among them a so-called `perfect' code using a minimum
number of only 5 qubits \index{quantum codes!perfect code}
\cite{laflamme:96,bennett-divincenzo-et-al:96}.  One can also
construct codes that are able to correct more than a single qubit
error. These satisfy a similar set of orthogonality relations of the
form given above, and the code words are entangled states of an
increasing number of qubits. An introduction to the theory of quantum
error correction can be found in the articles by Steane
\cite{steane-review:98,steane-SPRINGER:00} and by Gottesman
\cite{gottesman:00}, for example.

In quantum error correction we exploit, as in quantum algorithms, the
possibility to manipulate superpositions of states of a quantum
register and to measure joint observables which describe joint
properties of several qubits.  The operation $ENC^{-1}$ for example,
followed by one-qubit measurements, corresponds to the measurement of
the ``parity'' of different qubits, as was shown by Gottesman
\cite{gottesman:96}.  The joint observables are here\footnote{These
  observables generate an Abelian group, the so-called {\em
    \HAind{stabilizer}} of the Shor code \cite{gottesman:96}.}
\begin{eqnarray}
 M_1 &=& \sigma_{z,1}\sigma_{z,2}\,, \nonumber\\
 M_2 &=& \sigma_{z,2}\sigma_{z,3}\,,\nonumber\\ 
 M_3 &=& \sigma_{z,4}\sigma_{z,5}\,,\nonumber\\
 M_4 &=& \sigma_{z,5}\sigma_{z,6}\,,\nonumber\\
 M_5 &=& \sigma_{z,7}\sigma_{z,8}\,,\nonumber\\
 M_6 &=& \sigma_{z,8}\sigma_{z,9}\,,\nonumber\\ 
 M_7 &=& \sigma_{x,1}\sigma_{x,2}\sigma_{x,3}
  \sigma_{x,4}\sigma_{x,5}\sigma_{x,6}\,,\nonumber\\ 
 M_8 &=& \sigma_{x,4}\sigma_{x,5}\sigma_{x,6}
  \sigma_{x,7}\sigma_{x,8}\sigma_{x,9}\,.
\end{eqnarray}
The error spaces $\sigma_{\mu,j}{\cal H}_S$ are eigenspaces of these
observables with eigenvalues $\pm 1$. The observable
$M_1=\sigma_{z,1}\sigma_{z,2}$, for example, tells us whether on qubit
1 or 2 a spin flip has occurred without revealing any information on
which of the qubits: $M_1 (|000\rangle + |111\rangle) = + (|000\rangle
+ |111\rangle)$, $M_1 (|100\rangle + |011\rangle) = - (|100\rangle +
|011\rangle)$, $M_1 (|010\rangle + |101\rangle) = - (|010\rangle +
|101\rangle)$.  By measuring both observables
$M_1=\sigma_{z,1}\sigma_{z,2}$ and $M_2=\sigma_{z,2}\sigma_{z,3}$ one
can find out whether, and  
on which of the qubits 1,
2 , 3 a spin flip has taken place.

The measurement of the eigenvalues of these joint observables can be
realized, as described in Fig.~\ref{FIGqecc}, by the method ``decode
and subsequently measure the individual state of the surrounding
qubits''. This strategy has the disadvantage that the decoding leaves
the logical qubit in an unprotected state, exposing it directly to the
influence of decoherence.  There are different methods (or networks,
respectively) which use so-called ancillas to perform the error
detection and correction on the encoded state directly. This is the
subject of fault-tolerant quantum error correction and, more
generally, fault-tolerant quantum computation\index{quantum computation!fault-tolerant}. It takes into account
the fact that the elementary operations which are part of the encoding
and decoding network may themselves be imperfect and subject to
errors. Thus one needs to make sure that the correction operations do
not introduce more errors into the system than they extract.  A
description of the theory of fault-tolerant quantum computation is
beyond the scope of this introduction. A central result states that it
is possible, by using concatenated encoding strategies, to maintain the
coherence of the logical state of a quantum computer over an
arbitrarily long time, given that the error probability (noise level)
of the elementary operations (quantum gate, measurement) is below a
certain {\em threshold}\index{threshold!for fault-tolerant quantum computation}.  The price one has to pay for this is a
certain {\em overhead} in the number of auxiliary (physical) qubits
that scales polynomially [or under certain circumstances even
polylogarithmically] with the time of computation. The threshold is
very small and of the order of $10^{-4}-10^{-5}$. The theory of
fault-tolerant computation is considered as the general solution of
the problem of decoherence and imperfect apparatus for quantum
computation. An introduction into the subject is given by Preskill
\cite{preskill:98}, for example.

\bigskip

{\bf What is an Error?\index{error}}  Let us return to the question whether the
model of an error as a random unitary rotation is reasonable. The
interaction of the qubits with the environment can be described as a
unitary evolution in the Hilbert space of the total system consisting
of both the qubits and the environment.  Where and in what sense do
``errors'' happen in this picture?  This question is
certainly justified. One can show, however, that any interaction of
the qubits and the environment can be written in the (integrated) form
\begin{equation}
 |\phi\rangle_S|u\rangle_{\rm env} \rightarrow \sum_k 
(F_k|\phi\rangle_S) |u_k\rangle_{\rm env}
\label{digitalisation}
\end{equation} 
where the operators $F_k$ are tensor products of Pauli operators and
$|u_k\rangle_{\rm env}$ states of the environment which in general are
neither orthogonal nor normalized \cite{steane-SPRINGER:00}.  This
result remains true if $|\phi\rangle_S$ is replaced by an arbitrary
multi-qubit state \cite{steane-SPRINGER:00}.  In case of a quantum
error correcting code, such as the Shor code, one has the additional
property that $_S\langle \phi|F_k^{\rm 1 bit}|\phi\rangle_S = 0$, for
all ``1-bit operations'' $F_k^{\rm 1 bit}$ which contain a
non-trivial Pauli operator only at a single position (that is, for
$F_k^{\rm 1 bit} \sim \mathbb{I} \otimes \dots \otimes \mathbb{I}
\otimes \sigma_{\mu} \otimes \mathbb{I} \otimes \dots \otimes
\mathbb{I}$).  For weak interactions and for uncorrelated noise, these
are the terms of first order in an expansion with respect to the
interaction strength.  Since the overall evolution on the space of
the qubits plus the environment is unitary, the picture of randomly
occurring errors, which are subsequently corrected, is only a helpful
way of thinking about the problem.  The ``digitalization of noise''
\cite{steane-SPRINGER:00} is in fact only introduced via the
measurement of certain observables such as the $M_k$.  By measuring
the $M_k$ the state (\ref{digitalisation}) of the system is projected
``back'' into the code space ${\cal H}_S$ or any of the error spaces
$F_k^{\rm 1 bit} {\cal H}_S \bot {\cal H}_S$ that are orthogonal to
it.  It is the {\em dis}entanglement\index{disentanglement} of the qubits from the
environment which is the crucial process. The ``error'' is introduced
by the fact that one does not always project back to the code space
but sometimes also into an orthogonal error space, so that
subsequently a unitary correction operation has to be applied to
rotate the state back into the code space. On the Hilbert space of the
qubits alone, the entire process can effectively be described as if
the environment would apply random rotations $\sigma_{\mu,j}$ on the
code, which we then check and possibly correct.  Similar remarks apply
to codes that correct several errors at the same time, which take into
account terms of higher order in the expansion (\ref{digitalisation}).

\HAendIND{quantum error correction}



%% file: purification_ed.tex
\section{Entanglement Purification}
\label{sec:purificarion}
\HAbegIND{entanglement purification}

In \HAind{quantum communication}, entanglement between distant parties plays a
predominant role. In the following, we will concentrate on
communication scenarios which involve two parties, \HAind{Alice} and \HAind{Bob}. What
does it mean when we say that Alice and Bob have entanglement at their
disposal? Usually, this means that they own quantum systems whose
state is entangled, or, in technical terms, that the density operator
which describes the state of the two quantum systems cannot be written
as a convex combination of product states \cite{werner:89}. The two
entangled quantum systems are usually called \HAind{EPR pairs}, due to the
famous paper by Einstein, Podolsky and Rosen \cite{EPR:35}. In the
context of quantum information theory, the EPR pairs often consist of
two entangled two-level systems (qubits), one owned by Alice, and the
other by Bob. Maximally entangled two-qubit states are called
\emph{\HAind{Bell states}}; one can find four orthogonal Bell states, which
form a basis of the two-qubit Hilbert space, the \emph{\HAind{Bell basis}}.

The importance of entanglement is due to the fact that it is a
resource which is equivalent to a quantum channel: If Alice and Bob
are connected with a quantum channel, Alice can create an EPR pair
locally and send one half through the quantum channel to Bob. On the
other hand, if Alice and Bob own EPR pairs, they can use them to
teleport qubits \cite{bennett-brassard-et-al:93}, even when
they are not connected via some ``real'' quantum channel like an
optical fiber. 

The questions remains however, how can Alice and Bob obtain perfect
EPR pairs if they can only communicate via a noisy channel?  Any real
quantum channel interacts with the quantum systems which are
sent through it: it becomes entangled with them. %
This fact is important 
if Alice uses the channel in order to distribute EPR
pairs. If the EPR pairs are subsequently used for teleportation, then the
teleported qubits become entangled with the quantum channel.

\emph{Entanglement purification protocols}
\cite{BBPSSW:96,bennett-divincenzo-et-al:96,DEJMPS:96} can be 
used to overcome this problem. Simply speaking, these protocols create
an ensemble of highly entangled pairs out of a larger ensemble of pairs
with low \HAind{fidelity}. The fidelity of a quantum state $\rho$
is defined as its overlap with a given Bell state $\Ket{\Phi^+}$, say, \ie
\(F = \Bra{\Phi^+}\rho\Ket{\Phi^+}\).

The purified pairs provide Alice and Bob with a purified quantum
teleportation channel. If this channel is used for quantum
communication, the qubits are protected against an unwanted
interaction with the channel. In the next sections, we will see that
this fact can be exploited for quantum cryptography protocols.

In order to perform an entanglement purification protocol, classical
communication between Alice and Bob is necessary. This means, that
both Alice and Bob perform measurements on their respective qubits,
and tell each other the measurement outcomes. For some protocols only
\emph{one-way} communication is required, i.\,e.~only Alice will send
classical messages to Bob. It has been shown by Bennett \emph{et al.}
\cite{bennett-divincenzo-et-al:96}, that these one-way entanglement
purification protocols are equivalent to quantum error correcting
codes (see Section \ref{sec:quantum_error_correction}). 

A tutorial introduction to the basic idea of entanglement purification
is given in Ref.~\cite{briegel-SPRINGER:2000}

\subsection{2-Way Entanglement Purification Protocols}
\label{sec:2way_prot}

The two-way entanglement purification protocols (2-EPP) which we
present here have been developed by Bennett \emph{et al.}
\cite{BBPSSW:96} and, later, by Deutsch \emph {et al.}
\cite{DEJMPS:96}.  Since these protocols work in recursive way, they
are often referred to as \emph{recurrence protocols}.  In order to
distinguish between both protocols, we will call them
\emph{IBM}\index{entanglement purification!IBM protocol} and
\emph{Oxford}\index{entanglement purification!Oxford protocol} protocol, respectively. The IBM protocol introduces a
\emph{twirling} operation after each purification step, which
transforms the state of the EPR pairs into the Werner form. Since
Werner states \cite{werner:89} are described by only one real parameter, all
calculations can be done analytically.  A disadvantage of the IBM
protocol is that it is less efficient in producing pure states from
noisy ones than the Oxford protocol.  Qualitatively, there is no
difference between both protocols.

\begin{figure}[tbp]
  \centering
  \includegraphics[width=\figurewidth]{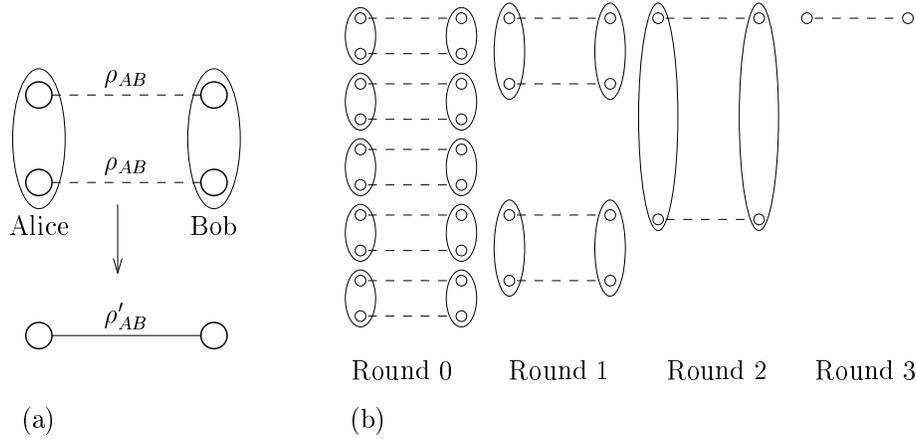}
  \caption[The entanglement purification protocol and the entanglement
  purification process.]{(a) The entanglement purification protocol
    \index{entanglement purification!protocol and process} is
    a (probabilistic) protocol, which creates a higher entangled pair
    of qubits out of two pairs with lower entanglement. Usually these
    pairs are called source and target pair, respectively. Through an
    interaction between the qubits of the source and the target pair,
    realizing a so-called \HAind{CNOT} operation on each side, the states of
    all four qubits become correlated. By measuring the qubits of the
    target pair, the source pair is probabilistically projected into a
    new state $\rho'_{AB}$, which is more entangled than the original
    state $\rho_{AB}$.%
    (b) The distillation process consists of several \emph{rounds}. In
    each round, the pairs are combined into groups of two at a time, and the
    purification protocol is applied to them. From round to round, the
    entanglement of the remaining pairs is increased. }
  \label{fig:prot_and_proc}
\end{figure}

To be precise, we want to distinguish between the purification
\emph{protocol} and the \emph{distillation process} (see Fig.
\ref{fig:prot_and_proc}).

In each step, the purification protocol acts on two pairs of qubits.
For the sake of simplicity, we shall assume that these two pairs are
described by the density operator $\rho_{AB}\otimes\rho_{AB}$, which
is thus a four-qubit density operator. The Oxford protocol (see
Fig.~{\ref{fig:prot_and_proc}) consist of the following steps:

\begin{enumerate}
\item Alice and Bob perform one-qubit $\pi/4$ rotations about the
  $x$-axis on each of their qubits (in opposite directions). If the
  qubits were stored in atomic/ionic degrees of freedom inside a trap,
  this could be implemented by (simple) laser pulses.
\item Both Alice and Bob perform a CNOT-operation (controlled NOT)
  \cite{springer:00},
  where they use their respective particle of pair one (two) as the
  source (target). This is the part of the protocol which is most
  difficult to perform experimentally.
\item Finally, both Alice and Bob measure the qubits which belong to
  pair two in the $\sigma_z$-basis, and tell each other the results
  (two-way communication). Whenever the results coincide, the keep
  pair one, otherwise they discard it. In either case, they have to
  discard the second pair, because it is projected onto a product
  state by the measurement.
\end{enumerate}

In order to see how this protocol works, it is useful to write the
density matrices in the \HAind{Bell basis}, i.\,e. in the basis of the two
qubit Hilbert space, which consists of the four Bell states
$\Ket{\Phi^\pm} = 1/\sqrt{2}\left(\Ket{00}\pm\Ket{11}\right)$ and
$\Ket{\Psi^\pm} = 1/\sqrt{2}\left(\Ket{01}\pm\Ket{10}\right)$:
\begin{equation}
  \label{eq:bell_diagonal}
  \rho_{AB} = A\Proj{\Phi^+} + B\Proj{\Psi^-} + C\Proj{\Psi^+} +
  D\Proj{\Phi^-} + \mathrm{off-diag.\atop elements}  
\end{equation}
The coefficients $A,B,C,$ and $D$ are called the \emph{Bell diagonal
  elements} of the density matrix $\rho_{AB}$. For any physical state,
these coefficients have to fulfill the normalization condition
$\tr{\rho_{AB}} = A+B+C+D = 1$. 

As it turns out, the Bell diagonal elements $A', B', C'$ and $D'$ of
the remaining pair do not depend on the off-diagonal elements of
$\rho_{AB}$. For this reason, we can find a recurrence relation for
the Bell diagonal elements, which describes their evolution during the
distillation process (the index $n$ belongs to the state of the pairs
at the beginning of round number $n$ in the distillation process:
\begin{equation}
  \label{eq:qpa_recursion}
  \begin{split}
    A_{n+1} &= \frac{A_n^2 + B_n^2}{N},\quad  B_{n+1} =
    \frac{2C_nD_n}{N}\\ 
    C_{n+1} &= \frac{C_n^2 + D_n^2}{N},\quad  D_{n+1} =
    \frac{2A_nB_n}{N}\\ 
  \end{split}
\end{equation}
The normalization $N_n = (A_n+B_n)^2 + (C_n+D_n)^2$ is equal to the
probability $p_\mathrm{success}$, that Alice and Bob obtain the same
measurement
results in step 3 of the protocol. Even though no analytical
solution has been found for this recurrence relation, it has been
shown (numerically in \cite{DEJMPS:96} and later analytically
\cite{macciavello98}) that it converges to the fixpoint
\index{entanglement purification!fixpoint} \(A^\infty=1,
B^\infty=C^\infty=D^\infty=0\), whenever the initial fidelity is
greater than $1/2$. In this case, also the off-diagonal elements will
vanish, since the density matrix has to be positive. In other words,
whenever Alice and Bob are supplied with EPR pairs with a fidelity of
more than 50\%, they can distill (asymptotically) perfect EPR pairs.

For the IBM protocol, one only needs one recurrence relation, since
(one-parametric) Werner states, described by $A = F, B=C=D = (1-F)/3$
and vanishing off-diagonal elements in (\ref{eq:bell_diagonal}), are
mapped onto Werner states. This map is shown in Fig.
(\ref{fig:bennett_puri_curve}a). The map has tree fixpoints. Two of
these fixpoints are attractive (at $F = 1/4$ and $F=1$), and the
remaining one (at $F=1/2$) is repulsive. Thus, if one starts the
distillation process with a fidelity greater than 1/2, one will
finally reach EPR pairs in a pure state. If the initial fidelity is
smaller than 1/2, one will finally be left with completely depolarized
pairs, which correspond to a Werner state with a fidelity of 1/4.
\begin{figure}[htbp]
  \centering
  \includegraphics[scale=0.8]{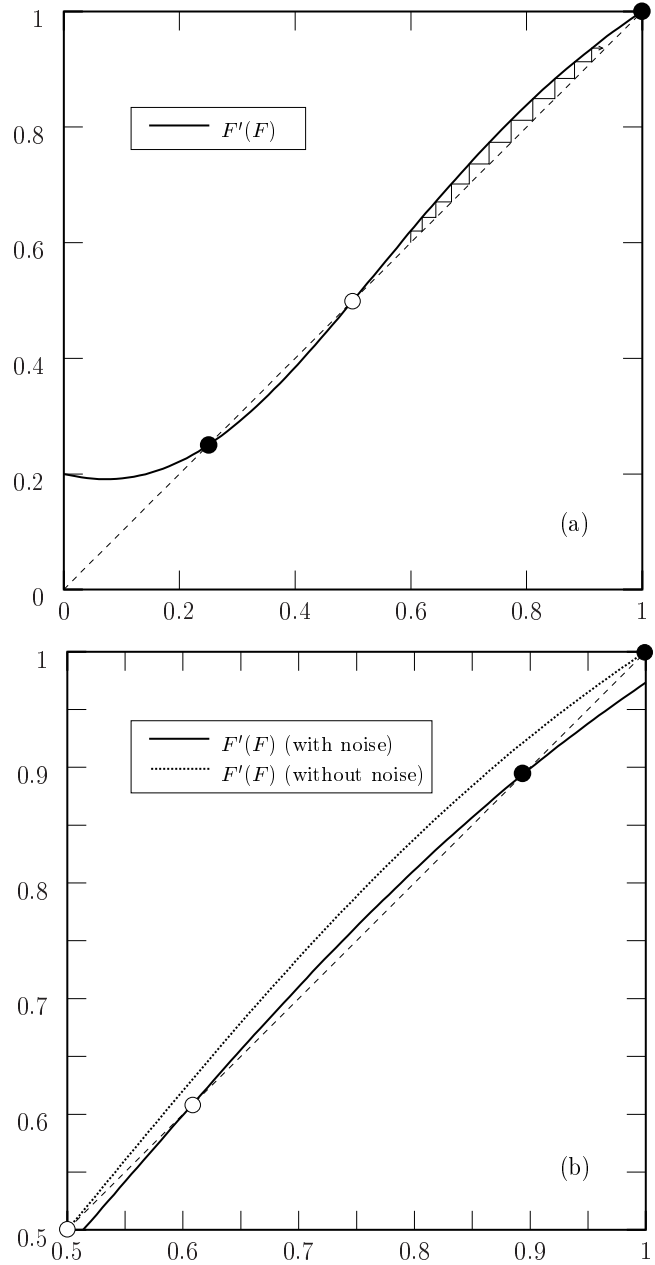}
  \caption[The purification curve for the IBM protocol]
    {The \HAind{purification curve} for the IBM protocol 
    \cite{BBPSSW:96,bennett-divincenzo-et-al:96} for perfect (i.\,e.
    noiseless) apparatus (a). The staircase denotes how the fidelity
    increases from round to round in the distillation process of
    Fig.~\ref{fig:bennett_puri_curve}b). If the apparatus is
    imperfect, the purification curve is ``pulled down'' (b) and the
    fixpoints move towards each other. The upper fixpoint of the
    curves indicates the maximum achievable fidelity $F_\mathrm{max}$,
    which can be reached asymptotically by the respective purification
    protocols; $F_\mathrm{max}$ decreases with an increasing noise
    level.  Attractive fixpoints are denoted by black circles,
    repulsive fixpoints by white circles.  }
  \label{fig:bennett_puri_curve}
\end{figure}


\subsection{Purification with Imperfect Apparatus}
\label{sec:noisy_purification}

Up to now, we have assumed that the only source of decoherence is the
quantum channel which connects Alice and Bob. For practical
implementations, however, this is an over-simplification. Indeed,
there are many operations involved in the distillation process: Qubits
have to be stored for a certain time, one- and two-qubit unitary
operations will act on them, and there are measurements. Each of these
operations is a source of noise by itself. It would be inconsistent to
ignore this source of noise. So the following question arises: What
are the conditions which we have to impose on the apparatus so that
entanglement distillation works at all?

As we have mentioned in the context of fault tolerant quantum
computation, there exists a certain noise threshold for the elementary
operations, below which fault tolerant quantum computation is
possible.  In the case of 2-EPP we will find a threshold
\index{threshold!for entanglement purification protocols}
 which is
much more favorable than the threshold for fault tolerant quantum
computation.

In order to get a qualitative understanding of the influence of noisy
operation on the entanglement distillation process, we look again at
the purification curve (Fig. (\ref{fig:bennett_puri_curve})). The
curve shows how the fidelity after a purification step depends on the
previous fidelity. If noise is introduced in the purification process
itself, it is intuitively clear that only a smaller increase in
fidelity can be achieved: the purification curve is ``pulled
down''.
In Fig. (\ref{fig:bennett_puri_curve}b) this is shown
schematically. We thus expect that in the case of noisy operations, one
has to start with a greater initial fidelity in order to purify at
all, and that the maximum fidelity which can be reached will be
smaller than unity.

If the noise level is increased, one reaches the situation that two of
the fixpoints will merge. 
At even higher noise levels, the
purification curve has only the trivial fixpoint which corresponds to
completely depolarized pairs: the distillation process breaks down and
does not work any longer.

The quantitative investigation of entanglement purification with 
\HAind{noisy apparatus}
\cite{giedke,briegel} shows that the above considerations are
qualitatively correct. For the calculation, the following noise model
has been assumed \cite{briegel}:

\begin{itemize}
\item The unitary evolution of the qubits is accompanied by a
  depolarizing channel. \index{quantum channel!depolarizing channel}
  It is well-known that this can be written in
  a time-integrated form
  \begin{equation}
    \label{eq:white_noise}
    \rho_{AB} \rightarrow p\,\, U_A \rho_{AB} U_A^{-1} +
    \frac{1-p}{d}\Eins_A \otimes \tr_A \rho_{AB}.
  \end{equation}
  Here, $\rho_{AB}$ is the density operator which describes the state
  of a bipartite quantum system, $U_A$ is the desired unitary
  operation (which is assumed to act only on the quantum system at
  party A), $d$ is the dimension of the Hilbert space of A's system,
  and $p$ is the \emph{\HAind{reliability}} of the quantum operation. For
  $p=1$, there is no noise at all, and for $p=0$, the quantum system
  at A becomes completely depolarized.
\item Measurements give the correct results only with a certain
  probability $\eta$. This can be conveniently described in terms of a
  \HAind{POVM} (positive operator valued measure \cite{peres:93}),
  \begin{equation}
    \label{eq:povm}
    \begin{split}
      M_0 & = \eta\Proj{0} + (1-\eta)\Proj{1}\\
      M_1 & = \eta\Proj{1} + (1-\eta)\Proj{0},
    \end{split}
  \end{equation}
  for one-qubit measurements in the $\sigma_z$ basis. Here,
  $\tr(M_j \rho)$ describes the probability with which the
  detector indicates the result ``$j$'' for the measured qubit.
\end{itemize}

As one can see from Eq.~(\ref{eq:white_noise}), we have to distinguish
between one- and two-qubit operations, if they are accompanied by
noise: a two-qubit depolarizing channel is different from two
one-qubit depolarizing channels. The first is an example of a
\emph{correlated} noise channel, the latter of an \emph{uncorrelated}
noise channel. The reliability of one- and two-qubit operations is
referred to as $p_1$ and $p_2$, respectively. Whether or not
entanglement purification is possible with a certain protocol, depends
on the three parameters $p_1,p_2$, and $\eta$. For all these parameters,
one gets a noise threshold
\index{threshold!for entanglement purification protocols}
 in the percent regime, which is about two
orders of magnitude better than the noise threshold for fault tolerant
quantum computation.

\HAendIND{entanglement purification}


%% file: cryptography_ed.tex
\section{Quantum Cryptography}
\HAbegIND{quantum cryptography}

\label{sec:cryptography}

One of the practically most advanced fields in quantum communication
is quantum cryptography. In this section, we will describe the two
basic protocols of quantum cryptography. We show that decoherence in
the (untrusted) quantum channel as well as in the (trusted) apparatus
plays an important role in the security analysis of quantum
cryptography protocols.

The communication scenario in the cryptographic context looks as
follows: Alice wants to send a confidential message
(\emph{clear-text}) to Bob, while a third communication party, Eve,
wants to listen in and learn as much as possible about the message. In
order to achieve her goal, Alice encrypts the message using some
cryptographic method. The encrypted message is called
\emph{ciphertext}.  A cryptographic protocol is considered
\emph{good}, if it is possible to restrict the information which Eve
can obtain to any desired level.

There exist several categories of \emph{classical} cryptographic
protocols; these include symmetric key ciphers, asymmetric key ciphers
and one-time pads. All these protocols have advantages and
disadvantages, but the most eminent advantage of the one-time pad is
that it has been proved to be secure in the information theoretical
sense: one can show that an eavesdropper can gain no information (zero
bits of information) about the message, even if he or she knows every
single bit of the encrypted message. To this end, it is however
necessary that Alice and Bob share a secret and random key, which
must at least be as long as the message which Alice wants to
transmit, and that this key will only be used once (thus the name
\emph{one-time pad}).

The one-time pad works as follows: As a key, Alice and Bob share a
secret string of zeros and ones \(s = (s_1, s_2, \ldots, s_N)\).
Similarly, Alice can write the clear-text (like any piece of
information) as a string of zeros and ones, using some encoding which
Alice and Bob agree on publicly. The clear-text is thus given in a
\emph{binary representation} \(t = (t_1, t_2, \ldots, t_N)\). For the
ciphertext, Alice adds the key and the clear-text bitwise modulo 2: \(c
= (s_1 \oplus t_1, s_2 \oplus t_2, \ldots, s_N \oplus t_N)\). In order
to decrypt the message, Bob simply adds the key bitwise (modulo 2) to
the ciphertext, and gets back the binary representation of the
clear-text.

The key used in the one-time pad protocol is a valuable resource, to
both the legitimate communication parties and to an eavesdropper:
Alice and Bob use up the key during the communication. In order to
supply themselves with a new key, they have to meet each other
physically. On the other hand, if \HAind{Eve} knows the key, the
communication 
between Alice and Bob is no longer a secret for her; for this reason,
the cryptographic key might be a valuable target for theft or bribery.
The aim of quantum cryptography is to solve this shortcoming of
classical cryptography. In most quantum cryptography protocols, the
quantum part of the protocol is related to the distribution of a key
(\HAind{quantum key distribution, QKD}), which can afterwards, as soon
as it 
is established, be used for a classical \HAind{one-time pad} protocol.

\subsection{The BB84 Protocol}
\label{sec:BB84}

The first protocol for quantum key distribution was given by Bennett
and Brassard in 1984 \cite{BB84}. This so-called BB84
protocol 
\index{quantum cryptography!BB84 protocol}
is widely used in quantum cryptography, since all security
considerations are well analyzed, and it is easy to understand.

The protocol works as follows: Alice prepares two random binary
strings, the key string \((k_1, k_2, \ldots, k_N)\) and the basis
string \((b_1, b_2, \ldots, b_N)\).  The randomness of the bits is
crucial for the security of the protocol; they may thus not be
chosen by a pseudo random number generator. 

There are 4 different quantum states which Alice can prepare:
\(\Ket{s_{00}} = \Ket{0}, \Ket{s_{01}} = \Ket{1}, \Ket{s_{10}} =
\Ket{+} \equiv 1/\sqrt{2}(\Ket{0}+\Ket{1}), \Ket{s_{11}} = \Ket{-}
\equiv 1/\sqrt{2}(\Ket{0}-\Ket{1})\). For simplicity, we will now
consider the case of qubits which are represented in the polarization
degree of freedom of a photon. In this case, the four states which
Alice can prepare are horizontally, vertically, or \(\pm 45^\circ\)
polarized photons. 

Alice sends $N$ photons through the quantum channel
to Bob. The state in which the qubits are prepared depends on the key-
and and the basis string: the $i$th qubit is prepared in the state
\(\Ket{s_{b_i k_i}}\).

Bob can measure each photon that arrives in his laboratory either in
the \(\Ket{0}/\Ket{1}\)-basis (\ie in the horizontal/vertical basis),
or in the \(\Ket{+}/\Ket{-}\)-basis (\ie in the \(\pm 45^\circ\)
polarized basis). For each individual photon, he selects the
measurement basis randomly, and he writes down the chosen basis and the
measurement result. When Bob has received and measured the $N$
photons, he is left with two strings of $N$ bits: the ``basis'' string and
the ``result'' string.

Alice and Bob exchange their respective basis strings through a
classical channel, which may be public; for example, they might
announce the basis strings in a newspaper. It is no security breach if
Eve knows both basis strings. However, Alice and Bob must make sure
that Eve cannot alter these messages. One possibility to achieve this
goal is that Alice and Bob posses an initial shared secret, which can
be used to check the authenticity and integrity of the basis strings.
During the key distribution task, this initial shared secret can be
recreated, so that it is not used up; rather, it plays the role of a
catalyst. By comparing their basis strings, Alice and Bob can see
which photons have been measured in the same basis in which they have
been prepared. Whenever the preparation basis and the measurement
basis are different, Bob's measurement result is completely random and
cannot be used. On the other hand, if the two bases are the same,
Bob's measurement result will be strictly correlated with Alice's key
bit for the respective photon: Alice's key bits and Bob's measurement
results for these photons can be used as a secret key.

Before the key can be used, Alice and Bob have to make sure that the
quantum channel has not been eavesdropped. One way to do this is the
following: Alice chooses a certain number of the key bits randomly and
sends them to Bob through the classical public channel. Bob compares
Alice's key bits with his result bits, and if they are equal, they can
be sure that there was no eavesdropper who tapped the quantum
channel. This is due to the fact that the only quantum operation which
does not disturb non-orthogonal quantum states is the identity. In
other words: if Eve does not want to disturb the non-orthogonal
quantum states which Alice sends, she has to leave them alone.

\subsection{The Ekert Protocol}
\label{sec:E91}

The main difference between the BB84 protocol and the so-called E91
protocol 
\index{quantum cryptography!E91 protocol}
found by Ekert in 1991 \cite{ekert:91} is that it does not
use single photons which one communication party sends to the other,
but pairs of entangled photons. While its experimental realization is
more difficult than the BB84 protocol, it has a theoretical advantage:
the security of the E91 protocol is related to the fact that there
exists no local realistic theory which explains the outcomes of
Bell-type experiments.\footnote{For a recent review of experiments
  testing \HAind{Bell's inequalities}, see e.\,g. \cite{zeilinger:99}.}  While
in the BB84 protocol one has to believe that the quantum mechanical
description of photons is complete (\ie that there exist no (local)
variables --- ``hidden'' or not --- which could be used to predict
Bob's measurement outcomes\footnote{In experiments, classical
  information about the state which has been prepared might leak out
  of Alice's laboratory through different degrees of freedom, like the
  frequency of the photon, or the polarization of a second photon in a
  multi-photon pulse. This information could \emph{in principle} be
  exploited by Eve without introducing noise. For the E91 protocol,
  this leakage problem does not exist, since such information does not
  exist until Alice and Bob perform their measurement.}), the E91
protocol performs a Bell experiment at the same time, which assures
that there cannot exist (local) hidden variables.

\begin{figure}[htbp]
  \centering
  \includegraphics[width=\figurewidth]{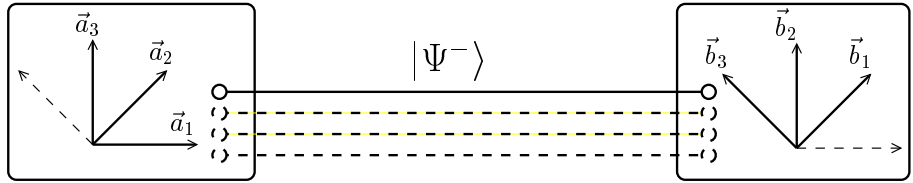}
  \caption{The measurement directions in the Ekert protocol. For each
    EPR pair, Alice and Bob choose independently and randomly one of
    the three measurement directions $\vec{a_1},\vec{a_2},\vec{a_3}$
    and $\vec{b_1},\vec{b_2},\vec{b_3}$, respectively.}
  \label{fig:ekert}
\end{figure}

In the E91 protocol, pairs of entangled photons are prepared, for
example in the state \(\Ket{\Psi^-}=(\Ket{01}-\Ket{10})/\sqrt{2}\). It
does not matter whether these pairs are produced in Alice's or Bob's
laboratory, or by a (potentially untrusted) source in between. One
photon of each pair is sent to Alice, the other to Bob. For each
photon, Alice and Bob choose one out of a set of three measurement
directions at random, and measure the polarization of the photon in
this direction (see Fig.~\ref{fig:ekert}).  
As in the BB84 protocol,
Eve must not be able to predict the choice of the measurement
directions. As soon as all pairs are sent to Alice and Bob and they
acknowledge that they have performed the measurements, the information
about the measurement directions is exchanged (through a public
classical channel). Alice and Bob check for which pairs their
respective measurement directions were the same; for all pairs where
they have chosen different measurement directions, also the
measurement outcomes are exchanged through the public classical
channel. With these results, Alice and Bob check that the Bell
inequalities \cite{bell:64,CHSH:69}
are violated. The measurement results for the pairs where
they have chosen the same measurement direction are strictly
anti-correlated, and can be used as a key.

\subsection{Security Proofs}
\label{sec:sec_proofs}

As we have seen above, the quantum key distribution protocols allow
for secure communication, as long as Alice and Bob are connected by a
noiseless quantum channel. This is a remarkable result -- however, it
would be useless for all practical purposes, since all quantum channels are
a source of noise. Since Alice and Bob trust only the equipment in
their laboratories, they cannot be sure that the noise which they
measure can be attributed to the channel. It is \emph{in principle}
impossible to distinguish between noise introduced by the quantum
channel or by an \HAind{eavesdropper}. For this reason, the communication
parties have to deal with the worst case scenario of an eavesdropper,
who is present all the time and everywhere, except  for the
laboratories, which are secure by assumption. The eavesdropper might be
hidden behind the noise of the quantum channel, and she might gain
partial knowledge of the cryptographic key and, later, of the secret
message.

The simplest way to deal with this situation would be to use a better quantum
channel. In a practical setting, however, when Alice and Bob are
connected by a given quantum channel (e.\,g. an optical fiber), this
possibility is ruled out. In this situation, Alice and Bob can use
\emph{privacy amplification} methods, where a shorter and perfectly secure
key is distilled out of a longer key, about which Eve might have had
considerable knowledge.  So-called ``ultimate'' or ``unconditional''
security proofs of quantum cryptography show that such protocols do exist.

The first of these proofs has been given by Mayers in 1996
\cite{mayers} for the BB84 protocol. Shor and Preskill gave a physical
interpretation of this proof, as they showed that it could \emph{a
  posteriori} be understood as a restricted, albeit sufficient, form
of quantum error correction and one-way entanglement purification.

A different approach has been taken by Deutsch \emph{et al.} in 1996
\cite{DEJMPS:96}.
They employ a two-way entanglement purification protocol (2-EPP, see
Sec.~\ref{sec:purificarion}) in order to distill almost pure EPR pairs
out of many imperfect pairs. If the purified pairs are used for
teleportation, the resulting quantum channel is perfectly secure:
Since the EPR pairs are in a pure state, they cannot be entangled with
any other quantum system. The eavesdropper is thus ``factored out'' in
the total Hilbert space, which we write symbolically as
\[\rho_\mathrm{Alice,Bob,Eve} \stackrel{ \mbox{
    \small2-EPP}}{\longrightarrow} \ProjInd{\Psi^+}{AB} \otimes
\rho_\mathrm{Eve}.\]

As we have already seen in Sec.~\ref{sec:noisy_purification}, in a
realistic setting the purification protocol does not converge to
perfect EPR pairs, but to some more or less mixed state in the Hilbert
space of Alice's and Bob's qubits. But that
means that the argument given above does no longer guarantee that Eve
is factored out: \emph{a priori}, there could exist residual
entanglement with Eve.\index{factorization of Eve}

\HAendIND{quantum cryptography}


%% file: factorization_ed.tex

\section{Private Entanglement}
\label{sec:priv_entanglement}

In the last section we have seen that entanglement purification (using
noisy apparatus) does not \emph{per se} guarantee a provably private
communication channel. Nevertheless, in this section we will show that
it suffices for the creation of ``\HAind{private entanglement}'', \ie
imperfect EPR pairs which are not entangled with an eavesdropper.
Private entanglement can thereby serve as noisy but secure quantum
channel.

The general idea is the following. Since Alice and Bob use noisy
apparatuses for the entanglement distillation process, it is clear that
the pairs become entangled with some degree of freedom of the
laboratory.  However, we will see that the total state of the
laboratory and of the (distilled) pairs converges to a pure state, and
then the same argument holds as in the case of noiseless entanglement
purification: a quantum system in a pure state cannot be entangled
with any other quantum system. In particular, Eve cannot be entangled
with the distilled pairs. These pairs can then be used for secure, albeit
noisy quantum teleportation.

In our analysis it is necessary to keep track of the state of the
laboratory. This seems to be a difficult task, since the the details
of the structure of the laboratory are unknown and complicated.  For
this reason one does usually not take care of these details, and
describes the system of qubits on which the noisy apparatus acts as an
open quantum system, with a master equation that describes their time
evolution \cite{gardiner:1991}. As an alternative, in the framework of
quantum information theory, we use the concept of completely positive
maps \cite{kraus_states}.

\subsection{The Lab Demon}
\label{sec:lab_demon}

\index{lab demon}
In this section, we give a simple model of a noisy
laboratory, which allows us to keep track of its state in terms of
classical variables.

As long as one cannot ``look into'' the device that introduced the
noise, there is no way to distinguish it from a different device whose
action is described by the same positive map. For this reason, our
simple noise model is sufficient for the proof, and we need not delve
into the complicated details of noisy quantum devices.

In order to keep the argument as transparent as possible, we will
restrict our attention to the case that only Alice's laboratory is a
source of noise; it would be easy to extend the argument to two noisy
laboratories.

Let us assume that in Alice's lab there is a little demon.  The lab
demon kicks and shakes the qubits from time to time, and is thus a
source of noise. However, there are no other sources of noise, and
even the lab demon acts on the qubits in a very controlled way: let us
assume that the demon has a random number generator that generates in
each time step pairs of numbers \((\mu,\nu) \in
\{0,1,2,3\}^{\times2}\), according to a given probability distribution
\(f_{\mu\nu}\) (which obeys the normalization condition
\(\sum_{\mu,\nu} f_{\mu\nu} = 1\)). The lab demon then applies the
(unitary) \HAind{error} operation \(\sigma_\mu^{(a_1)}\sigma_\nu^{(a_2)}\) to
the two qubits \(a_1\) and \(a_2\), on which Alice acts in the
entanglement purification protocol (see Section~\ref{sec:2way_prot}).
For \(\mu \in \{1,2,3\}\), the operators \(\sigma_\mu^{(a_i)}\) denote the
Pauli matrices acting on qubit $a_i$, and \(\sigma_0^{(a_i)} =
\Eins^{(a_i)}\). In addition, the lab demon writes down which error
operations he had applied to which qubits, since he will need this
informtion later.

Alice does not know which of the error operations have been applied to
the qubits, and she describes the action of the demon by the average
map
\begin{equation}
  \rho_{a_1a_2\ldots} \rightarrow  \sum_{\mu,\nu=0}^{3}
  f_{\mu\nu}\sigma_{\mu}^{(a_1)}\sigma_{\nu}^{(a_2)}
  \rho_{a_1a_2\ldots}\sigma_{\mu}^{(a_1)} 
  \sigma_{\nu}^{(a_2)}\,.
  \label{eq:noise_model}
\end{equation}
The ellipsis ($\ldots$) denotes other degrees of freedom, on which
Alice's lab demon does not act (like Bob's qubits, or some quantum
system in Eve's hands). We call the noise channel given by this
equation the \emph{correlated two qubit Pauli channel}. 
\index{quantum channel!correlated two qubit Pauli channel}
It includes,
for special choices of the probability distribution $f_{\mu\nu}$, the
one- and two-qubit depolarizing channel, and combinations thereof,
which have been studied in the context of entanglement purification
using imperfect apparatus in \cite{briegel}.

As mentioned above, we introduced the lab demon as a simplified noise
model in order to keep track of the internal state of the lab. For
that reason, we assume that the lab demon attaches an \emph{error
  flag}\index{error flag} $\lambda$ to each qubit. The error flag will represent four
different values, and it is convenient to divide it into two classical
bits. 

\subsection{The State of the Qubits Distributed Through the Channel}
\label{sec:state_after_channel}

\begin{figure}[htbp]
  \centering
  \includegraphics[width=0.45\figurewidth]{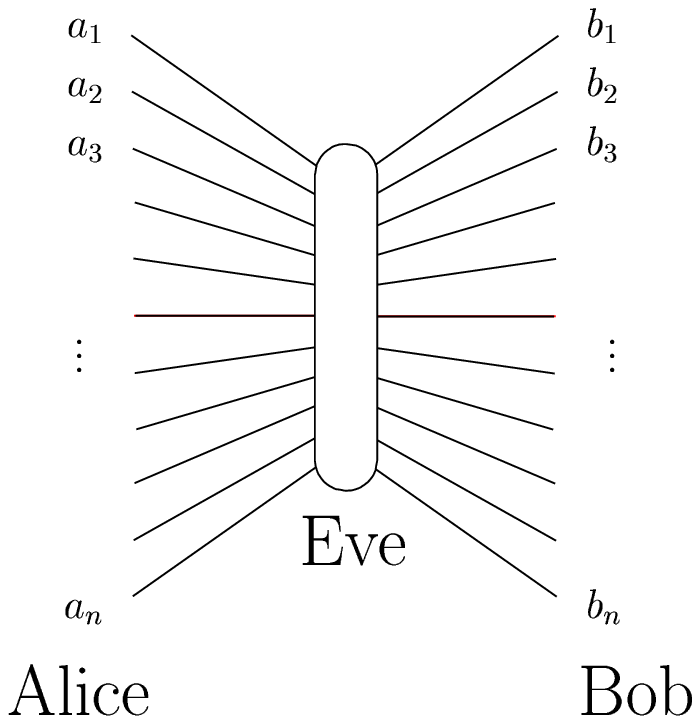}
  \hfill
  \includegraphics[width=0.45\figurewidth]{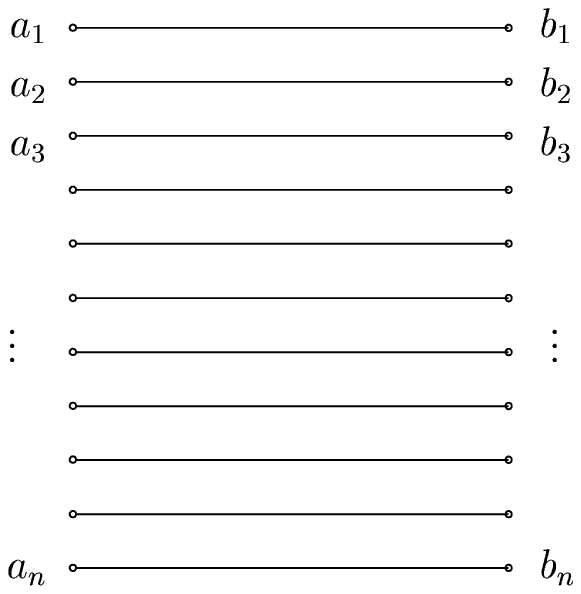}
  \hspace{0.25\figurewidth}(a)\hspace{0.5\figurewidth}(b)
  \caption{(a) The state of $N$ pairs which are distributed
    through the quantum channel is in the worst case scenario a
    general $2N$-qubit state, which might moreover be entangled with
    degrees of freedom under Eve's control. (b) After step 1, the
    state of the N pairs is a classically correlated ensemble of pure
    Bell states. 
  }
  \label{fig:spinne1}
\end{figure}
In the worst case scenario, all pairs which are distributed between
Alice and Bob have been prepared by Eve (see Fig.~\ref{fig:spinne1}a.
For that reason, the total state of all pairs is given by a general
$2N$-qubit density operator, which can be written in the form
\begin{equation}
\rho_{AB}=\BQuad\sum_{\mu_1\dots\mu_N \atop \mu_1' \dots\mu_N'} \BQuad
\alpha_{\mu_1\CLdots\mu_N \atop \mu_1' \CLdots\mu_N'} |{\cal
B}_{\mu_1}^{(a_1b_1)}\CCdots {\cal B}_{\mu_N}^{(a_Nb_N)}\rangle \!
\langle {\cal B}_{\mu_1'}^{(a_1b_1)}\CCdots {\cal
B}_{\mu_N'}^{(a_Nb_N)}|.
\label{rhoAB_1}
\end{equation}  
Here, $|\mathcal{B}_{\mu_j}^{(a_jb_j)}\rangle$, ${\mu_j}=00,01,10,11$
denote the 4 Bell states associated with the two qubits $a_j$ and
$b_j$ and \(j = 1,\ldots,N\).  Specifically, \(\Ket{\mathcal{B}_{00}}
\equiv \Ket{\Phi^+}\), \(\Ket{\mathcal{B}_{01}} \equiv \Ket{\Psi^+}\),
\(\Ket{\mathcal{B}_{10}} \equiv \Ket{\Phi^-}\), \(\Ket{\mathcal{B}_{11}}
\equiv \Ket{\Psi^-}\).

In general, (\ref{rhoAB_1}) will be an entangled $2N$-qubit state,
which might moreover be entangled with additional quantum systems in
Eve's hands. For the security analysis of the entanglement purification
protocol, this state is too complicated and cannot be handled. It
would be helpful if there was no entanglement between the different
pairs. Fortunately, Alice and Bob can apply the following protocol to the
pairs, in order to handle this situation:

\begin{description}
\item [Step 1:] On each pair of particles $(a_j,b_j)$, they apply
  randomly one of the four bi-lateral Pauli rotations
  $\sigma_k^{(a_j)}\otimes \sigma_k^{(b_j)}$, where k = 0,1,2,3.
\item [Step 2:] Alice and Bob randomly renumber the pairs,
  \((a_j,b_j)\to (a_{\pi(j)},b_{\pi(j)})\) where $\pi \in S(N)$ is a
  permutation which has been chosen at random.
\end{description}

It is important to note that Alice and Bob deliberately discard the
knowledge about which permutation and which of the Pauli rotations
have been applied to the pairs. Obviously, they cannot force Eve to do
the same thing.  So Eve might have a better description of the state
of the pairs than Alice and Bob. Thus the question remains whether
this additional knowledge might help Eve. It is easy to see that this
is not the case: Eve's description of the qubits has to be
\emph{statistically consistent} with the state which Alice and Bob or
the lab demon assign to the to qubits. As we are going to show, at the
end of the distillation process, the lab demon knows that the pairs
are pure EPR pairs. Eve can thus not have more information about the
pairs than the lab demon.\footnote{In fact, Eve has \emph{less}
  information than the lab demon, because she does not know the
  results of his random number generator.}

 After step 1, Alice's and Bob's
knowledge about the state is summarized by the density operator
\begin{equation}
  \tilde\rho_{AB}=\BQuad \sum_{\mu_1\dots\mu_N}\BQuad
  p_{\mu_1\dots\mu_N} |{\cal B}_{\mu_1}^{(a_1b_1)}\CCdots {\cal
    B}_{\mu_N}^{(a_Nb_N)}\rangle \langle {\cal B}_{\mu_1}^{(a_1b_1)}
  \CCdots {\cal B}_{\mu_N}^{(a_Nb_N)}|
\end{equation}    
which corresponds to a \emph{classically correlated ensemble} of pure
Bell states (see Fig.~\ref{fig:spinne1}b. The fact that the pairs are
classically correlated means that the order in which they appear in
the numbered ensemble may have some pattern, which may have been
imposed by Eve or by the channel itself. By applying step 2, the order
of the pairs is ``randomized''; this will prevent Eve from making use
of any possibly pre-arranged order of the pairs, which Alice and Bob
are meant to follow in the course of the distillation process: they
simply ignore this order.

The only correlation which remains is in the number of pairs which are
in a specific Bell state. In the limit of large $N$, it is consistent
for all relevant statistical predictions  to describe the ensemble with
the density operator

\begin{equation}
\label{rho_product}
\tilde{\tilde\rho}_{AB} = \left(\sum_{\mu}p_{\mu} |{\cal
B}_{\mu}\rangle \langle {\cal B}_{\mu}|\right)^{\otimes N} \equiv
(\rho_{ab})^{\otimes N}.
\end{equation} 

For finite $N$, the form of the state after step 2 is more
complicated; however, the subsequent arguments are also valid in that
case.

At this stage, Alice and Bob have to check whether the pairs are
``good enough'' for the distillation process, \ie they have to make sure
that the fidelity $F_0$ of the pairs is above the
purification/security threshold
\index{threshold!security threshold}
\index{threshold!for entanglement purification protocols}
(which coincide for all practical
purposes \cite{aschauer_pra}). 
They can do this by local measurements on a fraction of the
pairs and classical communication.

In order to separate conceptual from technical considerations and to
obtain analytical results, we will first concentrate on 
a toy model where all
the pairs are either in the state \(\Ket{\Phi^+}\) or
\(\Ket{\Psi^+}\). In this case, we talk about \emph{\HAind{binary pairs}}.

\subsection{Binary Pairs}
\label{binary_pairs}

Let us assume that Alice and Bob initially share pairs in the state 
\begin{equation}
  \label{eq:binary_pairs}
  \rho_{AB} = A \ProjInd{\Phi^+}{AB} + B \ProjInd{\Psi^+}{AB}
\end{equation}
(binary pairs) with $A = 1-B > 1/2$, and that the noise is of the form
(\ref{eq:noise_model}) with the restriction that the error
operators consist only of the identity and spin flip operators:

\begin{equation}
  \rho_{a_1a_2\ldots} \rightarrow  \sum_{\mu,\nu=0}^{1}
  f_{\mu\nu}\sigma_{\mu}^{(a_1)}\sigma_{\nu}^{(a_2)}
  \rho_{a_1a_2\ldots}\sigma_{\mu}^{(a_1)} 
  \sigma_{\nu}^{(a_2)}\,.
  \label{eq:binary_noise}
\end{equation}
Eq.~(\ref{eq:binary_noise}) describes a
\emph{two-bit correlated spin-flip channel}. The indices 1 and 2
indicate the source and target bit of the bilateral CNOT 
operation, respectively. It is straightforward to show that, using
this error model in the 2--EPP, binary pairs will be mapped onto
binary pairs.

At the beginning of the distillation process, Alice and Bob share an
ensemble of pairs described by (\ref{eq:binary_pairs}). 
In this special case, one bit suffices for the error flag.
 At this stage, all of these bits are set to zero. This reflects the
fact that the lab demon has the same \emph{a priori} knowledge about
the state of the ensemble as Alice and Bob.

In each purification step, two of the pairs are combined. The lab
demon first simulates the noise channel (\ref{eq:binary_noise}) on
each pair of pairs by using the random number generator as described.
Whenever he applies a \(\sigma_x\) operation to a qubit, he inverts
the error flag of the corresponding pair. Alice and Bob then apply the
2--EPP to each pair of pairs; if the measurement results in the last
step of the protocol coincide, the source pair will be kept.
Obviously, the error flag of that remaining pair will also depend on
the error flag of the the target pair, \ie the error flag of the
remaining pair is a function of the error flags of both ``parent''
pairs, which we call the \emph{\HAind{flag update function}}. In the case of
binary pairs, the flag update function maps two bits (the error flags
of \emph{both} parents) onto one bit. In total, there exist 16
different functions $g\!:\!\{0,1\}^2 \to \{0,1\}$. From these, the lab
demon chooses the logical AND function as the flag update function,
\ie the error flag of the remaining pair is set to ``1'' if and only
if both parent's error flags had the value ``1''.

After each purification step, the lab demon  divides all pairs into
two subensembles, according to the value of their error flags. By a 
straightforward calculation, we obtain for the coefficients $A_i$ and
$B_i$, which completely describe the state of the pairs in the
subensemble with error flag $i$, the following recurrence relations:

\begin{equation}
  \label{eq:recursion_binary}
  \begin{split}
    A_0' =  & \frac{1}{N}
    (f_{00}(A_0^2 + 2A_0A_1) + f_{11}(B_1^2+2B_0B_1)\\
    & +f_s(A_0B_1+A_1B_1+A_0B_0))\\
    A_1' =  & \frac{1}{N}
    \left(f_{00}A_1^2 + f_{11}B_0^2 + f_s A_1B_0 \right)\\
    B_0' =   &\frac{1}{N}
    (f_{00}(B_0^2 + 2B_0B_1) + f_{11}(A_1^2+2A_0A_1)\\
    &+f_s(B_0A_1+B_1A_1+B_0A_0))\\
    B_1' =  & \frac{1}{N}
    \left(f_{00}B_1^2 + f_{11}A_0^2 + f_s B_1A_0 \right)
  \end{split}
\end{equation}
with  \(N = (f_{00}+f_{11})((A_0+A_1)^2+(B_0+B_1)^2) + 2f_s (A_0+A_1)
(B_0+B_1) \) and \(f_s = f_{01} + f_{10}\). 

Since Alice and Bob do not know the values of the error flags, they
describe the pairs in terms of $A=A_0 + A_1$ and $B = B_0
+ B_1 = 1-A$ as in Eq.~(\ref{eq:binary_pairs}). The fidelity $F$ is thus
given by $F=A$.

For the case of uncorrelated noise, the error operations are applied
\emph{independently} and with probability $f_\mu (\mu = 0,1)$ to both qubits. This means that the probability
distribution \(f_{\mu\nu}\) factorizes into \(f_{\mu\nu} = f_\mu
f_\nu\). In this special case we obtain the following expression for
fixpoints of this map:
\begin{equation}
  \label{eq:binary_pairs_fixpoint}
  \begin{split}
    A_0^\infty & = \frac{1}{2} \pm \frac{\sqrt{f_0 - 3/4}}{f_0 - 1} 
    \quad\mathrm{or}\quad A_0^\infty = \frac{1}{2},\\
    A_1^\infty & = 0,\quad
    B_0^\infty   = 0 ,\quad
    B_1^\infty   = 1 - A_0^\infty  .  
  \end{split}
\end{equation}
The fixpoint of this map that is ``relevant'' for our discussion is
defined by the plus sign in the expression for \(A_0^\infty\) above.
It is not \emph{per se} clear that a fixpoint is also an attractor.
In fact, we find that Eq.~(\ref{eq:binary_pairs_fixpoint}) gives a
non-trivial fixpoint of (\ref{eq:recursion_binary}) for \(f_0 \ge
3/4\), but this fixpoint is an attractor only for
\(f_0 > f_0^\mathrm{crit}=0.77184451\) \cite{aschauer_pra}.

\begin{figure}[htbp]
  \begin{center}
    \includegraphics[width=\figurewidth]{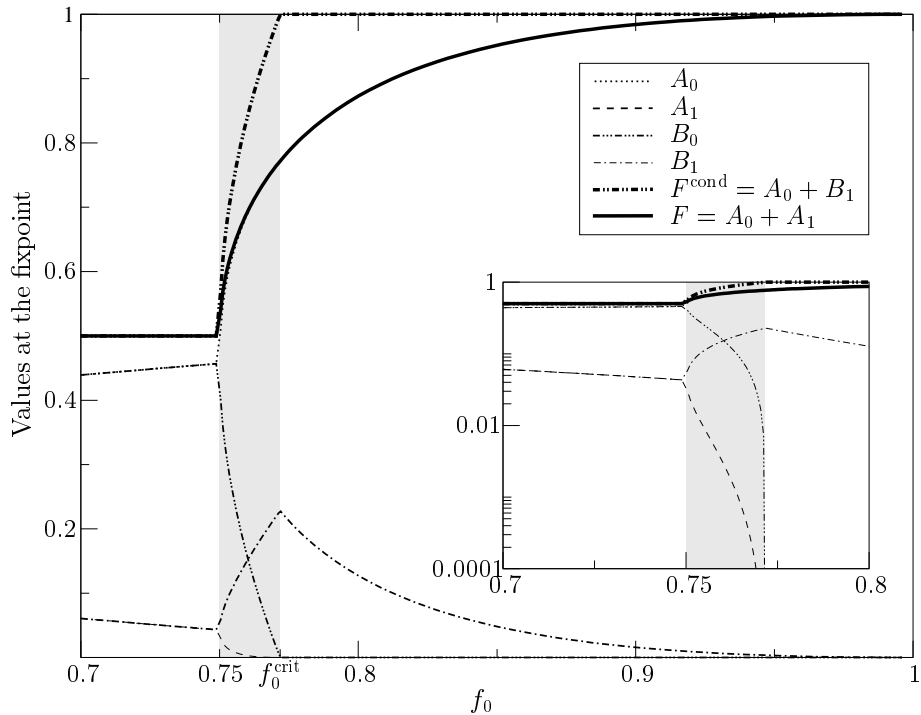}
    \caption[The values of \(A_0,A_1,B_0,B_1,F = A_0 +
    A_1,F^\mathrm{cond} = A_0 + B_1\) at the fixpoint as a function of
    the noise parameter \(f_0\).]
      {\label{fig:fixpoint_map}
      The values of \(A_0,A_1,B_0,B_1,F = A_0 + A_1,F^\mathrm{cond} =
      A_0 + B_1\) at the fixpoint as a function of the noise parameter
      \(f_0\) \cite{aschauer_pra}. For $f_0<0.75$, the values of $A_1$
      and $B_1$, as well as the values of $A_0$ and $B_0$ are equal,
      and the respective lines lie on top of each other.  One can
      clearly see that for $f_0 < 0.75$, the fidelity becomes $1/2$,
      and the pairs converge to the completely mixed state \(1/2
      \left( \Proj{\Psi^+} + \Proj{\Phi^+} \right)\): the protocol is
      not in the purification regime. For $f_0 > 0.75$, the maximum
      achievable fidelity increases, and approaches unity for $f_0 \to
      1$.  This corresponds to the fact that the protocol is in the
      purification regime, and that it works better if the apparatus
      is more reliable. However, the fidelity is strictly smaller than
      unity for $f_0 < 1$.  For the \HAind{conditional fidelity}
      \(F^\mathrm{cond} = A_0 + B_1\), however, the situation is
      different: above the critical value \(f_0^\mathrm{crit}\), it
      becomes strictly equal to unity. Since \(F^\mathrm{cond}\) is
      the fidelity or the pairs from the lab demon's point of view,
      any eavesdropper is factored out, and we call this regime the
      \emph{security regime}. The regime, where the protocol purifies
      but does not provide secure EPR pairs is called
      \emph{intermediate regime} (highlighted in grey).  The inset
      shows the same graphs on a logarithmic scale. In this graph, one
      can see that the parameters $A_1$ and $B_0$ do not vanish only
      asymptotically, but become zero at $f_0=f_0^\mathrm{crit}$.
      \index{entanglement purification!purification regime}
      \index{entanglement purification!security regime}
      \index{entanglement purification!intermediate regime}
    }
  \end{center}
\end{figure}

To summarize, we can identify three regimes for values of the noise
parameter (see Fig.~\ref{fig:fixpoint_map}): for a \emph{high} noise
level, when $f_0<3/4$, the protocol is not in the purification regime.
From Alice's and Bob's point of view, the protocol converges to the
completely mixed binary state \(1/2 \left( \Proj{\Psi^+} +
  \Proj{\Phi^+} \right)\). For a \emph{low} noise level, when $f_0 >
f_0^\mathrm{crit}$, the protocol converges to a state where \(A_0 +
B_1 = 1\) and \(A_1 = B_0 = 0\). This means that all pairs in the
subensemble 0 are in the state $\Ket{\Phi^+}$, and all pairs in the
subensemble 1 are in the state $\Ket{\Psi^+}$: From the lab demon's
point of view, all pairs are in a pure state! For that reason, we will
call this regime the \emph{security regime} of the entanglement
purification protocol. For \(3/4\le f_0 \le f_0^\mathrm{crit}\), the
protocol is in the \emph{intermediate regime}. This regime is of no
practical interest, since in this regime, the protocol converges very
slowly. However, the mere existence of the intermediate regime is
interesting, as it shows that purification and security are not
trivially related to each other.

As we have already seen, the sum \(A_0 + B_1\) is a measure for the
purity of these pairs from the lab demon's point of view. We call this
sum the \emph{\HAind{conditional fidelity}} $F^\mathrm{cond}$, since this is
the fidelity which Alice and Bob would assign to the pairs if they knew the values
of the error flags. 

\begin{figure}[htbp]
  \begin{center}
    \includegraphics{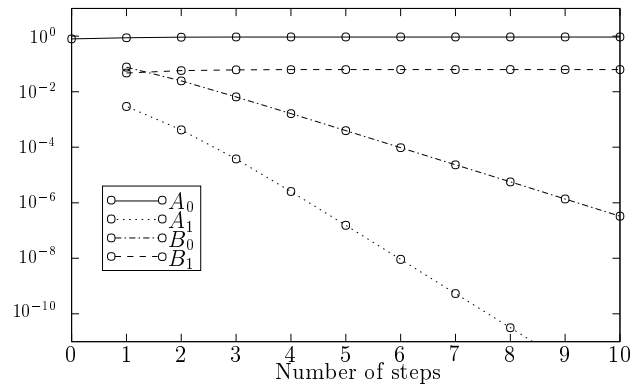}
    \caption{ \label{fig:binary_data}
      The evolution of the four parameters $A_0, A_1, B_0$, and $B_1$
      in the security regime. Note that both $A_1$ and $B_0$ decrease
      exponentially fast in the number of rounds. The initial fidelity
      was 80\%, and the values of the noise parameters were
      \(f_{00}=0.8575\), \(f_{01}=f_{10}=f_{11} = 0.0475\).}
  \end{center}
\end{figure}

We have also evaluated (\ref{eq:recursion_binary}) numerically in order to
investigate correlated noise (see Fig.~\ref{fig:binary_data}). Like in
the case of uncorrelated noise, we found that the coefficients $A_0$
and $B_1$ reach, during the distillation process, some finite value,
while the coefficients $A_1$ and $B_0$ decrease exponentially fast,
whenever the noise level is moderate.

\begin{figure}[htbp]
  \centering
  \includegraphics[width=\figurewidth]{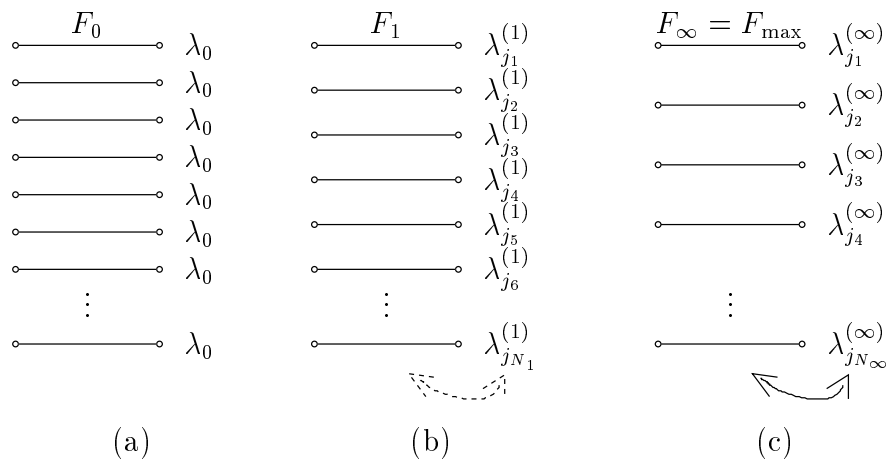}
  \caption{
    In the course of the distillation process, \emph{strict}
    correlations are built up between the state of the pairs and the
    error flags $\lambda_i$: (a) Initially, all error flags are set to
    zero, and there exist no correlations between the states of the
    pairs and the error flags; (b) after the first distillation round,
    there exist weak correlations. (c) Finally, in the
    asymptotic limit, the error flags are strictly correlated with the
    states of the pairs, and each flag identifies the state of the
    corresponding pair unambigously.
  }
  \label{fig:correlation}
\end{figure}

The distillation process which was described in
Fig.~\ref{fig:prot_and_proc}b now looks as in
Fig.~\ref{fig:correlation}, where the ensemble of pairs is now
supplemented with an error flag for each pair. One can see that in the
course of the distillation process, \emph{strict} correlations are
built up between the state of the pairs and the error flags
$\lambda_i$. In the asymptotic limit, each flag identifies the state
of the corresponding pair unambigously.

In other words, whenever the noise level is moderate, the conditional
fidelity converges to unity: entanglement purification can be used to
create private entanglement.

\subsection{Bell-Diagonal Initial States}

\index{Bell-diagonal states}
In the previous section we have considered the special case of binary
pairs. For arbitrary Bell diagonal states (Eq.~(\ref{rho_product}))
and for noise of the form (\ref{eq:noise_model}), the results are
quite similar. However, the most important difference is that in the
general case the intermediate regime is much smaller than in the case
of binary pairs. 

As already mentioned, in general the error flag consists of two
classical bits. This means that the the lab demon has to use a more
complicated flag update function than in the case of binary pairs. In
this case, the flag update function has been found by looking at how
errors are propagated during the course of the distillation process.
The details of this calculation can be found in \cite{aschauer_pra}.

Since the error flag represents four different values, the lab demon
divides all pairs into four subensembles, according to the value of
their error flag $\lambda$. In each of the subensembles the pairs are
described by a Bell diagonal density operator, like in
Eq.~(\ref{eq:bell_diagonal}), which now depends on the subensemble.
That means, in order to completely specify the state of all four
subensembles, we need 16 real numbers \(A^{ij},B^{ij},C^{ij}, D^{ij}\)
with \(i,j \in\{0,1\}\).

\begin{figure}[tp]
  \begin{center}
    \includegraphics[width=.3\figurewidth]{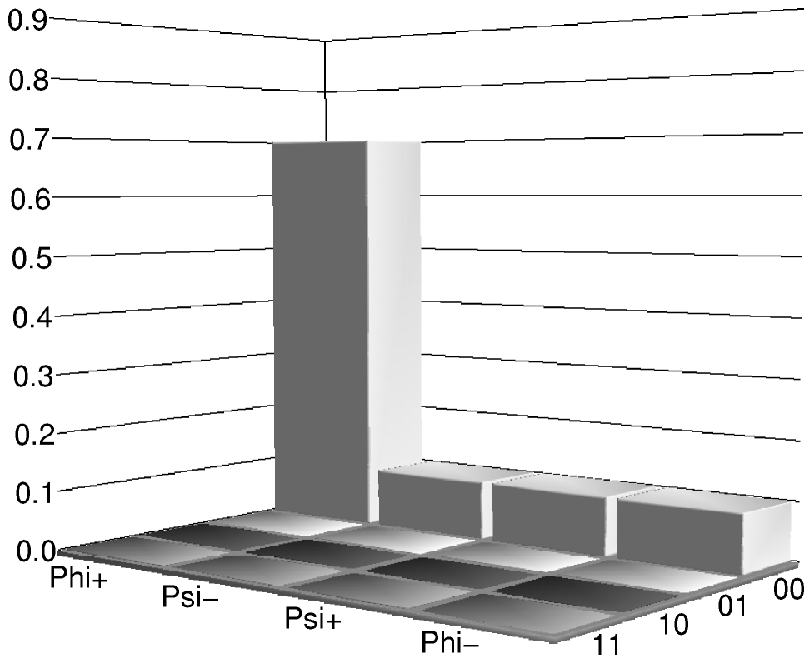}%
    \includegraphics[width=.3\figurewidth]{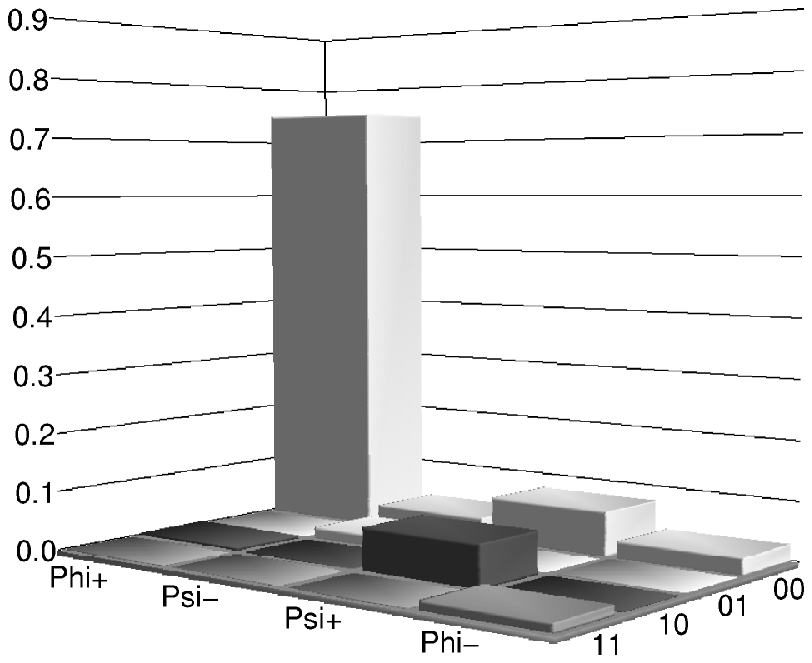}%
    \includegraphics[width=.3\figurewidth]{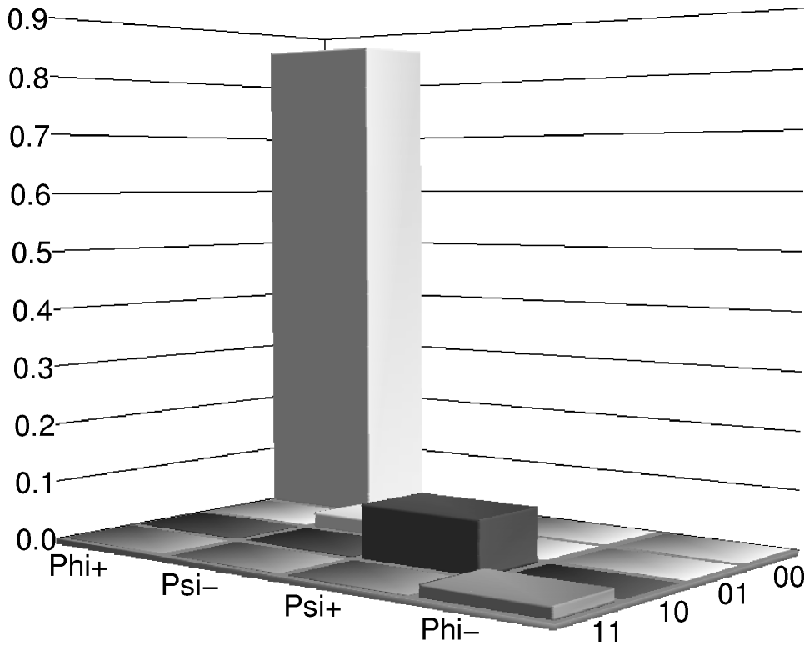}
    \caption{    
      \label{fig:state_evolution}
      Typical evolution of the 16 parameters \(A^{ij},B^{ij},C^{ij},
      D^{ij}\) with \(i,j \in\{0,1\}\) under the purification
      protocol.  As in Eq.~(\ref{eq:bell_diagonal}), the coefficients
      $A, B, C$, and $D$ correspond to the four Bell states
      which are indicated by ``Phi+'', ``Psi-'', ``Psi+'', and
      ``Phi-'' on one axis. The other axis shows the error flag
      \(\lambda \in \{00,01,10,11\}\). As one can see, only the
      diagoal elements survive, which means that the error flag
      identifies the states of all pair unambigously.  The noise
      parameters in this plot are $f_{00} = 0.83981, f_{0j} = f_{i0} =
      0.021131$ and $f_{ij} = 0.003712$ for $i,j \in \{1,2,3\}$.  }
  \end{center}
\end{figure}

Fig.~(\ref{fig:state_evolution}) shows how these 16 parameters evolve
under the action of the distillation process: If the protocol is in
the security regime, only the ``diagonal'' elements survive and are
identified by unambigously by the corresponding error flag. Again,
this means that from the lab demons point of view, all pairs are in a
pure state. 

\begin{figure}[htbp]
  \centering
  \includegraphics[width=\linewidth]{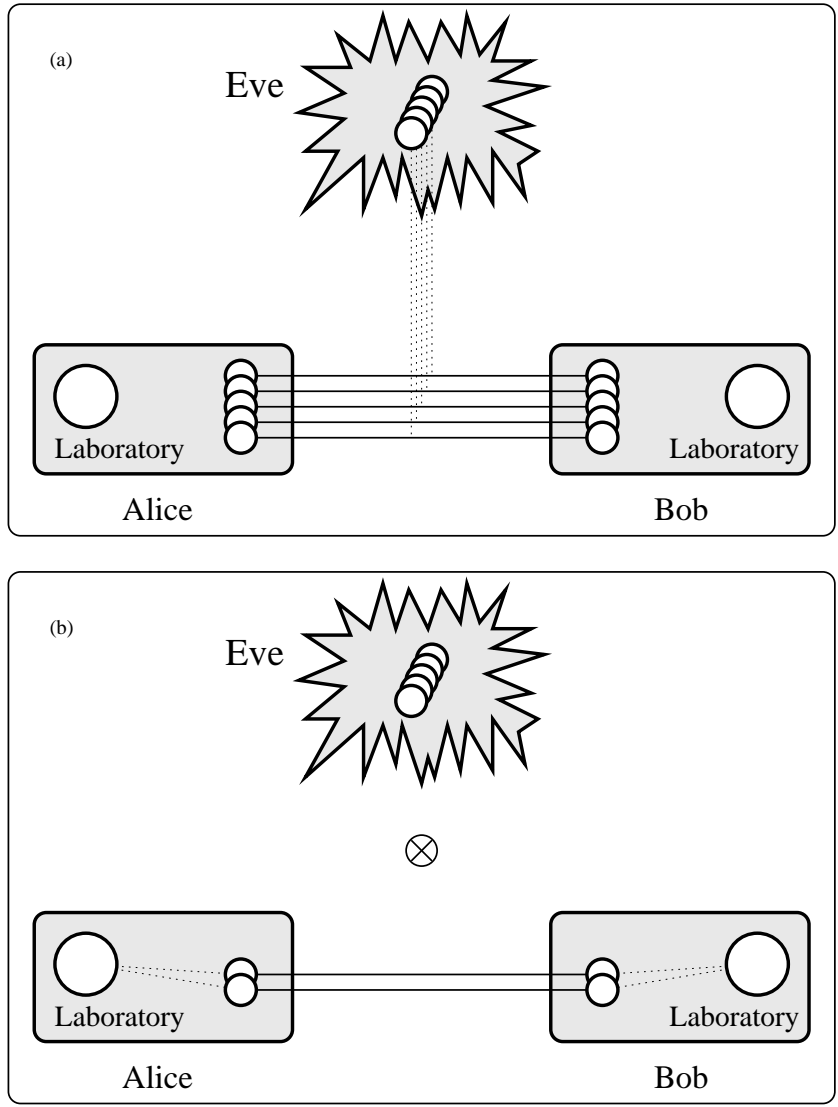}
  \caption{The entanglement distribution process \emph{redistributes}
    entanglement: at the beginning, the EPR pairs are entangled with
    the communication channel, or maybe even with an eavesdropper. In
    the end, however, the remaining EPR pairs are only entangled with
    the laboratories, and the eavesdropper is \emph{factored out}. }
  \label{fig:entanglement_flow}
\end{figure}

To summarize, we have found that in the entanglement distillation
process, entanglement is redistributed in the following sense (see
Fig.~\ref{fig:entanglement_flow}): in the beginning, there exists
(unwanted) entanglement between the EPR pairs and the quantum channel
(Eve). The entanglement distillation process is
not capable of creating perfect EPR pairs, since the pairs become
entangled with the laboratories, due to uncontrolled interactions.
Despite this fact, Eve is factored out, and all entanglement between her
and the EPR pairs is lost: Alice and Bob succeeded in creating private
entanglement and have thus a private, albeit noisy, quantum channel.

%% file: acknowledgement.tex
\subsection*{Acknowledgement}

We would like to thank the organizers of the
CohEvol workshop and seminar, who have managed
to bring together students and researchers
from different fields of physics in an inspiring
atmosphere.
This work has been supported by the Deutsche
Forschungsgemeinschaft (DFG) within the
Schwerpunktsprogramm QIV.
